\documentclass[twocolumn,amsmath,amssymb,floatfix,nofootinbib]{revtex4}

\usepackage{epsfig}
\usepackage{amsmath,amssymb}
\usepackage[all]{xy}

\newcommand{\beq}{\begin{equation}}
\newcommand{\eeq}{\end{equation}}
\newcommand{\lsi}{\,\raisebox{-0.13cm}{$\stackrel{\textstyle<}
{\textstyle\sim}$}\,}
\newcommand{\gsi}{\,\raisebox{-0.13cm}{$\stackrel{\textstyle>}
{\textstyle\sim}$}\,}
\newcommand{\be}{\begin{equation}}
\newcommand{\ee}{\end{equation}}

\begin{document}

\title{\bf Nuclear and nucleon transitions of the H di-baryon}
\author{Glennys R. Farrar and Gabrijela Zaharijas}
\affiliation{\it Center for Cosmology and Particle Physics\\
New York University, NY, NY 10003,USA}

\begin{abstract}
We consider 3 types of processes pertinent to the phenomenology of
an H di-baryon: conversion of two $\Lambda$'s in a doubly-strange
hypernucleus to an H, decay of the H to two baryons, and -- if the
H is light enough -- conversion of two nucleons in a nucleus to an
H. We compute the spatial wavefunction overlap using the
Isgur-Karl and Bethe-Goldstone wavefunctions, and treat the weak
interactions phenomenologically.  The observation of $\Lambda$
decays from doubly-strange hypernuclei puts a constraint on the H
wavefunction which is plausibly satisfied.  In this case the H is very long-lived as we calculate.  An absolutely stable H
is not excluded at present.  SuperK can provide valuable limits.
\end{abstract}

 \maketitle

%\vspace{4pt}
%\vfill\eject

\section{Introduction}

The  most symmetric color-spin representation of six quarks ($uuddss$) is called the H dibaryon. It is flavor singlet
with strangeness -2, charge 0, and spin-isospin-parity
$I(J^P)=0(0^+)$. In 1977 Jaffe calculated its mass \cite{jaffe} to be about 2150 MeV in the MIT-bag model and thus predicted it would be a strong-interaction-stable bound state, since decay to two $\Lambda$ particles would not be kinematically allowed. Since then its mass has been estimated in many different models, with results lying in the range 1 -- 2.3 GeV.  On the experimental side, there
have been many inconclusive or unsuccessful attempts to produce
and detect it.  See \cite{H} for a review.

The purpose of this paper is to study several processes involving
the H which are phenomenologically important if it exists:
conversion of two $\Lambda$'s in a doubly-strange hypernucleus to
an H, decay of the H to two baryons, and -- if the H is light
enough -- conversion of two nucleons in a nucleus to an H.  The
amplitudes for these processes depend on the spatial wavefunction
overlap of two baryons and an H. We are particularly interested in
the possibility that the H is tightly bound and that it has a mass
less than $m_N + m_\Lambda$.  In the case, as we shall see, its
lifetime is longer than the age of the Universe.

If the H is tightly bound, it would be expected to be spatially
compact.  Hadron sizes vary considerably, for a number of reasons.  For instance the nucleon is
more than twice as large as the pion, with charge radius $r_N =
0.88$ fm compared to $r_\pi = 0.38$ fm.  Lattice and instanton-liquid
studies can account for these diverse radii and further predict
that the glueball is even more tightly bound: $r_G \approx (1/4-1/6)
~r_N$ \cite{shuryak:RG}. If the analogy suggested in ref.
\cite{kf:Lam1405} between H, $\Lambda_{1405}$ and glueball is
correct, it would suggest $r_H \approx r_G \lsi 1/4 ~r_N$.  The above size relationships make sense:  the nucleon's large size is due to the low mass of the pion which forms an extended cloud around it, while the H and glueball do not couple to pions, due to parity and flavor conservation, are thus are small compared to the nucleon.   In the
absence of an unquenched, high-resolution lattice QCD calculation
capable of a reliable determination of the H mass and size, we
will consider all values of $m_H $ and take $r_H/r_N \equiv
1/f$ as a parameter, with $f$ in the range 2-6. For a more
detailed discussion of the motivation and properties of a stable
or long-lived H and a review of experimental constraints on such
an H, see ref. \cite{f:stableH}.

In this paper we calculate the lifetime for decay of the H to
various final states, and we consider two types of experimental
constraints on the transition of two baryons to an H in a nucleus,
$A_{BB}\rightarrow A'_{H}X$.   To estimate the rates for these
processes requires calculating the overlap of initial and final
quark wavefunctions.  We model that overlap using an Isgur-Karl
harmonic oscillator model for the baryons and H, and the
Bethe-Goldstone wavefunction for a nucleus. The results depend on
$r_N/r_H$ and the nuclear hard core radius.

Experiments observing single $\Lambda$ decays from double
$\Lambda$ hypernuclei $A_{\Lambda \Lambda}$\cite{ags,kek2}
indicate that $\tau(A_{\Lambda \Lambda}\rightarrow A_{H}'X) \gsi
 10^{-10}$ sec.  Our calculations show that adequate suppression of
$\Gamma(A_{\Lambda \Lambda}\rightarrow A_{H}'X)$ requires $r_H
\lsi 1/2~r_N$, consistent with expectations,  Thus an H with mass $m_H
< 2 m_\Lambda$ can still be viable in spite of the observation of
double-$\Lambda$ hypernuclei, as also found in ref. \cite{kahana}.

We calculate the lifetime of the H, in three qualitatively distinct mass ranges, under the assumption that the conditions to satisfy the  constraints from double-$\Lambda$ hypernuclei are met.  The ranges are $m_H < m_N + m_\Lambda$, in which H decay is a doubly-weak $\Delta S = 2$ process,  $m_N + m_\Lambda < m_H  < 2 m_\Lambda$, in which the H can decay by a normal weak interaction, and $m_H > 2 m_\Lambda$, in which the H is strong-interaction unstable.  The H lifetime, in these ranges respectively, is a few$\times 10^{12}$ yr, about a month, and $\sim 10^{-9}$ sec.

Finally, if $m_H \lsi 2 m_N$, nuclei are unstable to $\Delta S=-2$ weak
decays converting two nucleons to an H.  In this case, the
stability of nuclei is a more stringent constraint than the
double-$\Lambda$ hypernuclear observations, but our results show that nuclear stability bounds can also be satisfied if the H is
sufficiently compact: $r_H \lsi ~1/4 ~r_N$ depending on mass and
nuclear hard core radius, although this option is vulnerable to experimental exclusion by SuperK.

This paper is organized as follows. In section \ref{expts} we
describe in greater detail the two types of experimental
constraints on the conversion of baryons to an H in a nucleus.  In
section \ref{overlapcalc} we setup the theoretical apparatus to
calculate the wavefunction overlap between H and two baryons.  We
determine the weak interaction matrix elements phenomenologically
in section \ref{weakME}. Lifetimes for various
processes are computed in sections \ref{convlifetimes} and \ref{metastable}.
The results are reviewed and conclusions are summarized in section
\ref{summary}.

\section{Experimental constraints}
\label{expts}

\subsection{Double  $\Lambda $ hyper-nucleus detection}

There are five experiments which have reported positive results in
the search for single $\Lambda$ decays from double $\Lambda$
hypernuclei. We will describe them briefly. The three early
emulsion based experiments \cite{prowse,danysz,kek} suffer from
ambiguities in the particle identification, and therefore we do not consider them further. In the latest emulsion experiment at KEK
\cite{kek2}, an event has been observed which is
interpreted with good confidence as the sequential decay of ${\rm
He}^6 _{\Lambda \Lambda}$ emitted from a $\Xi ^-$ hyperon nuclear
capture at rest. The binding energy of the double $\Lambda$ system
is obtained in this experiment to be $B_{\Lambda \Lambda }=1.01\pm
0.2$ MeV, in significant disagreement with the results of previous
emulsion experiments, finding $B_{\Lambda \Lambda }\sim 4.5$ MeV.

The BNL experiment \cite{ags} used the $(K^-, K^+)$
reaction on a ${\rm Be}^9$ target to produce S=-2 nuclei. That
experiment detected pion pairs coming from the same vertex in the
Be target. Each pion in a pair indicates one unit of strangeness
change from the (presumably) di-$\Lambda$ system. Observed peaks in the two
pion spectrum have been interpreted as corresponding to
two kinds of decay events. The pion kinetic energies in those
peaks are (114,133) MeV and (104,114) MeV. The first peak can be
understood as two independent single $\Lambda$ decays from
$\Lambda \Lambda$ nuclei. The energies of the second peak do not
correspond to known single $\Lambda$ decay energies in
hyper-nuclei of interest. The proposed explanation\cite{ags} is
that they are pions from the decay of the double $\Lambda$ system,
through some specific He resonance. The required resonance has not
yet been observed experimentally, but its existence is considered
plausible. This experiment does not suffer from low statistics or
inherent ambiguities, and one of the measured peaks in the two
pion spectrum suggests observation of consecutive weak decays of a
double $\Lambda$ hyper-nucleus. The binding energy of the double
$\Lambda$ system $B_{\Lambda \Lambda }$ could not be determined in
this experiment.

The KEK and BNL experiments are generally accepted to demonstrate quite conclusively, in two
different techniques, the observation of $\Lambda$ decays from
double $\Lambda$ hypernuclei.  Therefore $\tau _{A_{\Lambda
\Lambda}\rightarrow A_{H}'X}$ cannot be much less than $\approx
10^{-10}$s.  (To give a more precise limit on $\tau _{A_{\Lambda
\Lambda}\rightarrow A_{H}'X}$ requires a detailed analysis by the
experimental teams, taking into account the number of hypernuclei
produced, the number of observed $\Lambda$ decays, the acceptance,
and so on.)  As will be seen below, this constraint is readily
satisfied if the H is compact: $r_H \lsi 1/2 ~r_N$.

\subsection{Stability of nuclei}

There are a number of possible reactions by which two nucleons can
convert to an H in a nucleus if that is
kinematically allowed ($m_H \lsi 2 m_N$).  The initial nucleons are most likely to be
$pn$ or $nn$ in a relative s-wave, because in other cases the
Coulomb barrier or relative orbital angular momentum suppresses
the overlap of the nucleons at short distances which is necessary
to produce the H. If $m_H \lsi 2 m_N - n
m_\pi$\footnote{Throughout, we use this shorthand for the more
precise inequality $m_H < m_{A} - m_{A'} - m_X$ where $m_X$ is the
minimum invariant mass of the final decay products.}, the final
state can be $H \pi^+ $ or $H \pi^0$ and $n-1$ pions with total
charge 0.  For $m_H \gsi 1740$ MeV, the most important reactions
are $p n \rightarrow H e^+ \nu_e$ or the radiative-doubly-weak
reaction $n n \rightarrow H \gamma$.

The best experiments to place a limit on the stability
of nuclei are proton decay experiments. Super Kamiokande (SuperK),
can place the most stringent constraint due to its large mass.  It is
a water Cerenkov detector with a 22.5 kiloton fiducial mass,
corresponding to $8~10^{32}$ oxygen nuclei. SuperK is sensitive to
proton decay events in over 40 proton decay
channels\cite{SuperK}. Since the signatures for the transition of
two nucleons to the H are substantially different from the
monitored transitions, a specific analysis by SuperK is needed to
place a limit.  We will discuss the order-of-magnitude of the
limits which can be anticipated.

Detection is easiest if the H is light enough to be produced with
a $\pi^+$ or $\pi^0$. The efficiency of SuperK to detect neutral
pions, in the energy range of interest (KE $\sim 0-300$ MeV), is
around 70 percent. In the case that a $\pi ^+$ is emitted, it can
charge exchange to a $\pi ^0$ within the detector, or be directly
detected as a non-showering muon-like particle with similar
efficiency.  More difficult is the most interesting mass range
$m_H \gsi 1740$ MeV, for which the dominant channel $p n
\rightarrow H e^+ \nu$ gives an electron with $E \sim (2 m_N -
m_H)/2 \lsi  70$ MeV.   The  channel
$nn \rightarrow H \gamma$, whose rate is smaller by a factor of order $\alpha$, would give a monochromatic
photon with energy $(2 m_N - m_H) \lsi 100$ MeV.

We can estimate SuperK's probable sensitivity as follows.  The
ultimate background comes primarily from atmospheric neutrino
interactions, $\nu N \rightarrow N'(e,\mu),\quad \nu N \rightarrow
N'(e,\mu)+n\pi $ and $ \nu N\rightarrow \nu N' +n\pi $, for which the event
rate is about 100 per kton-yr.  Without a
strikingly distinct signature, it would be difficult to detect a
signal rate significantly smaller than this, which would imply
SuperK might be able to achieve a sensitivity of order
$\tau_{A_{NN}\rightarrow A_{H}'X}\gsi {\rm few} 10^{29}$ yr. Since
the H production signature is not more favorable than the
signatures for proton decay, the SuperK limit on
$\tau_{A_{NN}\rightarrow A_{H}'X}$ can at best be $0.1 \tau_p$,
where $0.1$ is the ratio of Oxygen nuclei to protons in water.
Detailed study of the spectrum of the background is needed to make
a more precise statement. We can get a lower limit on the SuperK
lifetime limit by noting that the SuperK trigger rate is a few
Hz\cite{SuperK}, putting an immediate limit $\tau_{O\rightarrow H
+ X }\gsi {\rm few} 10^{25}$ yr, assuming the decays trigger
SuperK.

SuperK limits will apply to specific decay channels, but other
experiments potentially establish limits on the rate at which
nucleons in a nucleus convert to an H which are independent of
the H production reaction. These experiments place weaker
constraints on this rate due to their smaller size, but they are
of interest because in principle they measure the stability of
nuclei directly. Among those cited in ref. \cite{pdb}, only the
experiment by Flerov et. al.\cite{flerov} could in principle be
sensitive to transitions of two nucleons to the H. It searched for
decay products from ${\rm Th}^{232}$, above the Th natural decay
mode background of 4.7 MeV $\alpha$ particles, emitted at the rate
$\Gamma _{\alpha}=0.7~10^{-10} {\rm yr}^{-1}$. Cuts to remove the
severe background of 4.7 MeV $\alpha$'s may or may not remove
events with production of an H. Unfortunately ref. \cite{flerov}
does not discuss these cuts or the experimental sensitivity in
detail. An attempt to correspond with the experimental group, to
determine whether their results are applicable to the H, was
unsuccessful. If applicable, it would establish that the lifetime
$\tau_{{\rm Th}^{232}\rightarrow H + X}> 10^{21}$ yr.

Better channel independent limits on $N$ and $NN$
decays in nuclei have been established recently, as summarized in
ref. \cite{BOREXINO}. Among them, searches for the radioactive
decay of isotopes created as a result of $NN$ decays of a parent
nucleus yield the most stringent constraints.  This method was
first exploited in the DAMA liquid Xe detector \cite{DAMA}.
BOREXINO has recently improved these results\cite{BOREXINO} using
their prototype detector, the Counting Test Facility (CTF) with
parent nuclei ${\rm C}^{12},{\rm C}^{13}~{\rm and}~{\rm O}^{16}$.
The signal in these experiments is the beta and gamma radiation in
a specified energy range associated with deexcitation of a daughter
nucleus created by decay of outer-shell nucleons in the parent
nucleus. They obtain the limits $\tau _{pp} >
5~10^{25}$ yr and $\tau _{nn} > 4.9~10^{25}$ yr.  However H
production requires overlap of the nucleon wavefunctions at short
distances and is therefore suppressed for outer shell nucleons,
severely reducing the utility of these limits.  Since the SuperK limits
will probably be much better, we do not attempt to estimate the
degree of suppression at this time.

Another approach could be useful if for some reason the direct
SuperK search is foiled. Ref. \cite{suzuki} places a limit on the
lifetime of a bound neutron, $\tau _{n}>4.9~10^{26}$ yr, by
searching for $\gamma$'s with energy $E_{\gamma}=19-50$ MeV in the
Kamiokande detector. The idea is that after the decay of a neutron
in oxygen the de-excitation of $O^{15}$ proceeds by emission of
$\gamma$'s in the given energy range. The background is especially
low for $\gamma$'s of these energies, since atmospheric neutrino
events produce $\gamma$'s above 100 MeV. In our case, some of the
photons in the de-excitation process after conversion of $pn$ to
H, would be expected to fall in this energy window.

\section{Overlap of H and two baryons}
\label{overlapcalc}

We wish to calculate the amplitudes for a variety of processes,
some of which require one or more weak interactions to change
strange quarks into light quarks. By working in pole
approximation, we factor the problem into an H-baryon-baryon
wavefunction overlap times a weak interaction matrix element
between strange and non-strange baryons, which will be estimated
in the next section. For instance, the matrix element for the
transition of two nucleons in a nucleus $A$ to an H and nucleus
$A'$, $A_{NN} \rightarrow A'_H X $, is calculated in the $\Lambda
\Lambda$ pole approximation, as the product of matrix elements for
two subprocesses: a transition matrix element for formation of the
H from the $\Lambda \Lambda$ system in the nucleus, $ |{\cal M}|
_{\{\Lambda \Lambda\} \rightarrow H~X}$, times the amplitude for a
weak doubly-strangeness-changing transition, $|{\cal M}|_{NN
\rightarrow \Lambda \Lambda}$.  We ignore mass differences between
light and strange quarks and thus the spatial wavefunctions of all
octet baryons are the same.  In this section we are concerned with
the dynamics of the process and we suppress spin-flavor indices.

\subsection{Isgur-Karl Model and generalization to the H}
\label{IK}

The Isgur-Karl (IK) non-relativistic harmonic oscillator quark
model\cite{IK,faiman,bhaduri} was designed to reproduce the
masses of the observed resonances and it has proved to be
successful in calculating baryon decay rates \cite{faiman}. In the
IK model, the quarks in a baryon are described by the Hamiltonian
\beq \label{hamiltonian} H=\frac {1}{2m} (p^2 _1+p^2 _2+p^2 _3)
+\frac{1}{2}K\Sigma_{i<j} ^3 (\vec {r}_i -\vec {r}_j)^2, \eeq where
we have neglected constituent quark mass differences.  The wave
function of baryons can then be written in terms of the relative
positions of quarks and the center of mass motion is factored out.
The relative wave function in this model is \cite{faiman,bhaduri}
\beq \Psi _{B} (\vec{r}_1,\vec{r}_2,\vec{r}_3) = N_{B} \exp \left[
{-\frac {\alpha_{B} ^2}{6}\Sigma_{i<j} ^3 (\vec {r}_i -\vec
{r}_j)^2}\right], \eeq where $N_B$ is the normalization factor,
$\alpha _B=\frac {1}{\sqrt{<r_B ^2>}}=\sqrt{3Km}$, and $<r_B ^2>$
is the baryon mean charge radius squared. Changing variables to
\beq \label{rholambda} \vec {\rho} =\frac {\vec {r_1} -\vec
{r_2}}{\sqrt{2}},~\vec {\lambda}=\frac {\vec {r_1} +\vec {r_2}-2
\vec {r_3}}{\sqrt{6}} \eeq reduces the wave function to two
independent harmonic oscillators. In the ground state \beq
\Psi_{B} (\vec {\rho}, \vec {\lambda})=\left( \frac
{\alpha_B}{\sqrt{\pi}} \right) ^3 \exp\left[ -\frac
{\alpha_{B}^2}{2} (\rho ^2 + \lambda ^2)\right]. \eeq

One of the deficiencies of the IK model is that the value of the
$\alpha _B$ parameter needed to reproduce the mass splittings of
lowest lying $\frac {1}{2} ^+$ and $\frac {3}{2} ^+$ baryons,
$\alpha_B = 0.406 ~{\rm fm}^{-1}$, corresponds to a mean charge
radius squared for the proton of $<r^2_{ch}>= \frac {1}{\alpha _B
^2}=0.49$ fm. This is distinctly smaller than the experimental
value of 0.86 fm. Our results depend on the choice of $\alpha_B$
and therefore we also report results using $\alpha_B = 0.221~ {\rm
fm}^{-1}$ which reproduces the observed charge radius at the expense of the mass-splittings. 

Another concern is the
applicability of the non-relativistic IK model in describing quark
systems, especially in the case of the tightly bound H. With
$r_H/r_N = 1/f$, the quark momenta in the H are $\approx f$ times
higher than in the nucleon, which makes the non-relativistic
approach more questionable than in the case of nucleons.
Nevertheless we adopt the IK model because it offers a tractable
way of obtaining a quantitative estimate of the effect of the
small size of the H on the transition rate, and there is no other
alternative available at this time.

We fix the wave function for the H particle starting from the same
Hamiltonian (\ref{hamiltonian}), but generalized to a six quark
system.  For the relative motion part this gives \beq
\Psi_{H}=N_{H}\exp\left[-\frac{\alpha_{H}^2}{6}\sum _{i<j} ^6
(\vec{r_i} -\vec{r_j})^2\right]. \eeq The space part of the matrix
element of interest,\\ $<A'_{H}|A_{ \Lambda \Lambda }>$, is given by
the integral \beq \int \prod _{i=1} ^6 d^3\vec{r}_i \Psi
_{\Lambda} ^{a} (1,2,3) \Psi _{\Lambda} ^{b} (4,5,6) \Psi_H
(1,2,3,4,5,6). \eeq Therefore it is useful to choose variables for
the H wavefunction as follows, replacing \beq
\vec{r}_1,\vec{r}_2,\vec{r}_3,\vec{r}_4,\vec{r}_5,\vec{r}_6
\rightarrow \vec {\rho}^{a},\vec{\lambda}^{a},\vec{\rho}^{b},\vec
{\lambda}^{b}, \vec {a}, \vec {R}_{CM} \eeq where $\vec
{\rho}^{a(b)}$ and $\vec {\lambda}^{a(b)}$ are defined as in eq
(\ref{rholambda}), with $a(b)$ referring to coordinates
$1,2,3~(4,5,6)$.  (Since we are ignoring the flavor-spin part of
the wavefunction, we can consider the six quarks as
distinguishable and not worry about fermi statistics at this
stage.)  We also define the center-of-mass position and the
separation, $\vec {a}$, between initial baryons $a$ and $b$: \beq
\label{coord} \vec {R}_{CM}=\frac {\vec {R}_{CM}^{a}+\vec
{R}_{CM}^{b}}{2},~ \vec {a}=\vec {R}_{CM}^{a}-\vec {R}_{CM}^{b}.
\eeq Using these variables, the H ground state wave function
becomes
\begin{eqnarray}
\Psi_{H}&=&\left( \frac{3}{2}\right) ^{3/4}
\left( \frac{\alpha _H}{\sqrt{\pi}} \right)^{15/2}\\
&\times & \exp[-\frac {\alpha_{H} ^2}{2} (\vec {\rho^{a}}^2 + \vec
{\lambda ^{a}}^2+\vec {\rho^{b}}^2 + \vec {\lambda ^{b}}^2 +\frac
{3}{2} \vec {a}^2)]. \nonumber
\end{eqnarray}
As for the 3-quark system, $\alpha _H=\frac {1}{\sqrt{<r_H ^2>}}$.

\subsection{Bruecker-Bethe-Goldstone Nuclear Wavefunction} \label{BBG}

To describe two $\Lambda$'s or nucleons in a nucleus we use
solutions of the Bruecker-Bethe-Goldstone equation describing the
interaction of a pair of fermions in an independent pair
approximation; see, e.g., \cite{walecka}. The solution of the
Schrodinger equation for two fermions in the Fermi sea interacting
through a potential $v({\vec x}_1,{\vec x}_2)$ takes the form \beq
\psi (1,2)=\frac {1}{\sqrt{V}}~e^{i{\vec P}{\vec R}_{CM}}
%~\frac{1}{\sqrt{V}}
~\psi ({\vec a}) \eeq where ${\vec R}_{CM}$ and ${\vec a}$ are
defined as in (\ref {coord}). The first factor contains the
center-of-mass motion and the second is the internal wave function
of the interacting pair.  $\psi ({\vec a})$ is a solution of the
Bethe Goldstone equation (eq (36.15) in \cite{walecka}) which is
simply the Schrodinger equation for two interacting fermions in a
Fermi gas, where the Pauli principle forbids the appearance of
intermediate states that are already occupied by other fermions.
Both wave functions are normalized so that the space integral of
the wave function modulus squared equals one.
In the application of this equation to nuclear matter,
the interaction of each particle from the pair with all particles
in nuclei through an effective single particle potential is
included, in the independent pair approximation known as Bruecker
theory (see eq (41.1) and (41.5) in \cite{walecka}).

We are interested in s-wave solutions to the above equation since
they are the ones that penetrate to small relative distances.
Following \cite{walecka}, an s-wave solution of the internal wave
function is sought in the form \beq \psi (a)\sim \frac {u(a)}{a}
\eeq which simplifies the Bethe Goldstone equation to \beq ( \frac
{d^2}{dx^2}+k^2 )u(a)=v(a)u(a)-\int ^{\infty} _0\chi
(a,y)v(y)u(y)dy \eeq where $v(a)$ is the single particle potential
in the effective-mass approximation, and the kernel $\chi (a,y)$
is given by \beq \chi (a,y)=\frac {1}{\pi} \left[ \frac {\sin k_F
(a-y)}{a-y}-\frac {\sin k_F (a+y)}{a+y}\right]. \eeq

For the interaction potential between two nucleons in a nucleus we
choose a hard core potential for the following reasons.  The two
particle potential in a nucleus is poorly known at short
distances. Measurements (the observed deuteron form factors, the
sums of longitudinal response of light nuclei,...) only constrain
two-nucleon potentials and the wave functions they predict at
internucleon distances larger than $0.7$ fm \cite{pandharipande}.
The Bethe-Goldstone equation can be solved analytically when a
hard-core potential is used.  While the hard-core form is surely
only approximate, it is useful for our purposes because it enables
us to isolate the sensitivity of the results to the short-distance
behavior of the wavefunction. We stress again, that more
``realistic" wavefunctions are in fact not experimentally
constrained for distances below 0.7 fm. Rather, their form at
short distance is chosen for technical convenience or aesthetics.

Using the hard core potential, the s-wave BG wavefunction is \beq
\Psi_{BG}(\vec{a})=\left\{
\begin{array}{ll}
N_{BG}\frac{u(k_F a)}{k_F a} & \textrm{for \quad $a>\frac{c}{k_F}$} \\
0  & \textrm {for $\quad a<\frac{c}{k_F}$}
\end{array}\right.
\eeq where $\frac{c}{k_F}$ is the hard core radius.  Expressions
for $u$ can be found in \cite{walecka}, eq. (41.31). The
normalization factor $N_{BG}$ is fixed setting the integral of
$|\psi _{BG}|^2$ over the volume of the nucleus equal to one. The
function $u$ vanishes at the hard core surface by construction and
then rapidly approaches the unperturbed value, crossing over that
value at the so called ``healing distance''. At large relative
distances and when the size of the normalization volume is large
compared to the hard core radius, $u(a)/a$ approaches a plane wave
and the normalization factor $N_{BG}$ reduces to the value
$1/\sqrt{V}$, as \beq \psi _{BG}(a)=N_{BG}~\frac{u(k_F a)}{k_F
a}~\rightarrow \frac {1}{\sqrt{V}}~e^{ika}. \eeq

\subsection{Overlap Calculation}

The non-relativistic transition matrix element for a transition
$\Lambda \Lambda \rightarrow H$ inside a nucleus is given by
(suppressing spin and flavor)
\begin{eqnarray} \label{matrixel}
T_{\{\Lambda \Lambda\}\rightarrow H}&=&2 \pi i \delta (E) \int d^3
a~ d^3 R_{CM} \prod _{i=a,b}
d^3 \rho^i d^3 \lambda ^i \nonumber \\
&\times &  ~\psi^* _H \psi ^a _{\Lambda}~\psi^b _{\Lambda}~\psi
_{nuc}~ e^{i({\vec k}_H-{\vec k}_{\Lambda \Lambda}){\vec R}_{CM}} %\nonumber
\end{eqnarray}
where $\delta (E)=\delta (E_H-E_{\Lambda \Lambda})$, $\psi ^{a,b}
_{\Lambda}=\psi ^{a,b} _{\Lambda}(\vec {\rho}^{a,b},\vec
{\lambda}^{a,b})$, and  $\psi _{nuc}=\psi _{nuc}({\vec a})$ is the
relative wavefunction function of the two $\Lambda 's$ in the
nucleus.  The notation $\{\Lambda \Lambda\}$ is a reminder that
the $\Lambda$'s are in a nucleus. The plane waves of the external
particles contain normalization factors $1/\sqrt{V}$ and these
volume elements cancel with volume factors associated with the
final and initial phase space when calculating decay rates. The
integration over the center of mass position of the system gives a
3 dimensional momentum delta function and we can rewrite the
transition matrix element as \beq \label{matrixel2} T_{\{\Lambda
\Lambda\} \rightarrow H}=(2\pi)^4 i\delta ^4(k_f-k_i)~{\cal
M}_{\{\Lambda \Lambda\} \rightarrow H}, \eeq where $|{\cal
M}|_{\{\Lambda \Lambda\} \rightarrow H}$ is the integral over the
remaining internal coordinates in (\ref{matrixel}). In the case of
pion or lepton emission, plane waves of the emitted particles
should be included in the integrand. For brevity we use here the
zero momentum transfer, $\vec {k} =0$ approximation, which we have
checked holds with good accuracy; this is not surprising since
typical momenta are $\lsi 0.3$ GeV.

Inserting the IK and BBG wavefunctions and performing the Gaussian
integrals analytically, the overlap of the space wave functions
becomes
\begin{eqnarray} \label{overlap}
|{\cal M}|_{\Lambda \Lambda \rightarrow H}&=&\frac {1}{4} \left
(\frac {2f}{1+f^2}\right )^6 \left( \frac{3}{2}
\right)^{3/4}\left( \frac{\alpha _H}{\sqrt{\pi}} \right)^{3/2}\\
\nonumber &\times & N_{BG}\int^{\infty} _{\frac{c}{k_F}} d^3 a
\frac {u(k_F a)}{k_F a}e ^{-\frac {3}{4}\alpha_{H} ^2 a^2}
\end{eqnarray}
where the factor 1/4 comes from the probability that two nucleons
are in a relative s-wave, and $f$ is the previously-introduced
ratio of nucleon to H radius; $\alpha _H=f~\alpha _B $. Since
$N_{BG}$ has dimensions  $V^{-1/2}$ the spatial overlap ${\cal M}
_{\{\Lambda \Lambda\} \rightarrow H}$ is a dimensionless quantity,
characterized by the ratio $f$, the Isgur-Karl oscillator
parameter $\alpha_B$, and the value of the hard core radius.  Fig.
1 shows $|{\cal M}|^2_{\{\Lambda \Lambda\} \rightarrow H}$ versus
the hard-core radius, for a range of values of $f$, using the
standard value of $\alpha_B= 0.406~{\rm fm}^{-1}$ for the IK
model\cite{bhaduri} and also $\alpha_B = 0.221~{\rm fm}^{-1}$ for
comparison.

%\vspace{5mm}
%\centerline{\epsfig{file=ovlapGZ.eps,width=8cm}}{\footnotesize\textbf{Figure
%1:} Log$_{10}$ of $|{\cal M}|^2_{\Lambda \Lambda \rightarrow H}$
%versus hard core radius in fm, for ratio $f=R_N/R_H$ and two
%values of the Isgur-Karl oscillator potential.}
%\vspace{5mm}\\

\vspace{5mm}
\centerline{\epsfig{file=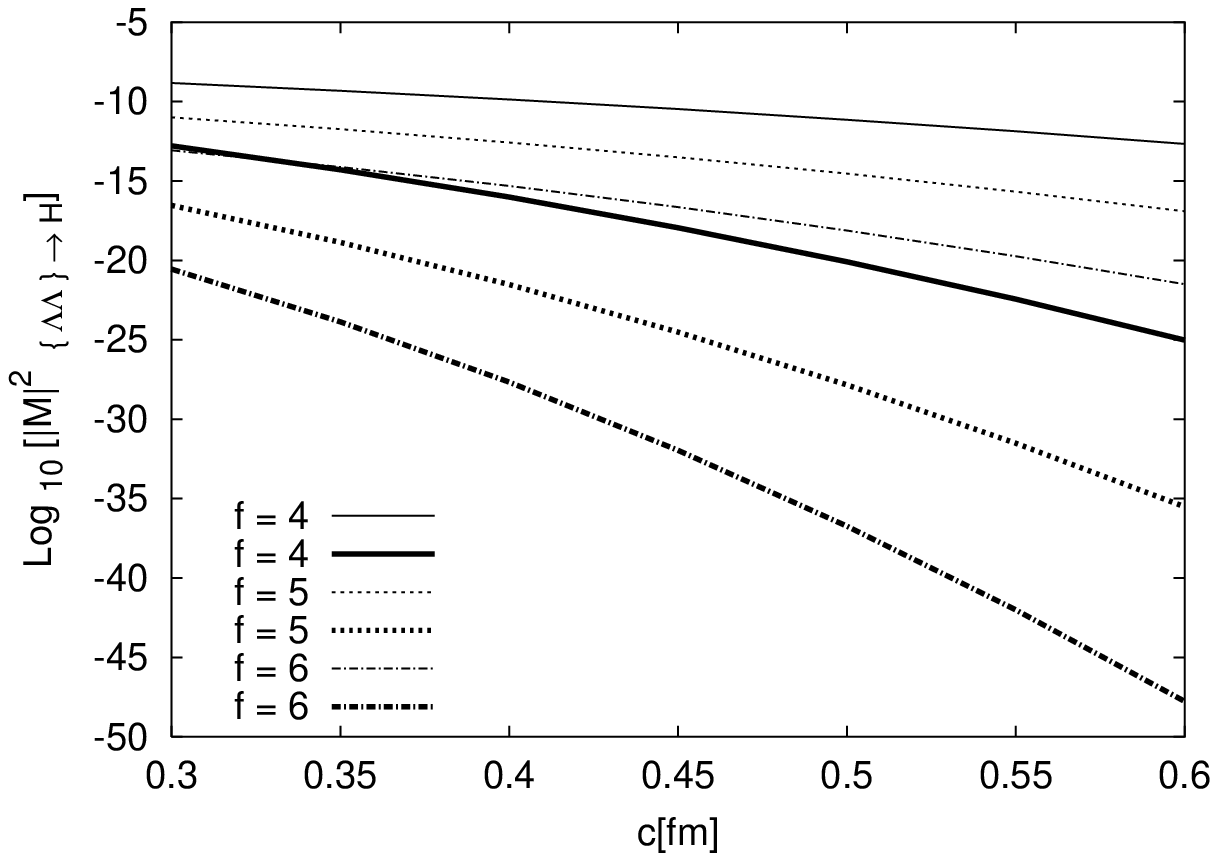,width=9cm}}{\footnotesize\textbf{Figure
1:} Log$_{10}$ of $|{\cal M}|^2_{\Lambda \Lambda \rightarrow H}$
versus hard core radius in fm, for ratio $f=R_N/R_H$ and two
values of the Isgur-Karl oscillator potential, with $\alpha_B= 0.406~{\rm fm}^{-1}$ (thick lines), 
and $\alpha_B = 0.221~{\rm fm}^{-1}$ (thin lines).}
\vspace{5mm}\\

\section{Weak Interaction Matrix Elements}
\label{weakME}

Transition of a two nucleon system to off-shell $\Lambda
\Lambda$ requires two strangeness changing weak reactions.
Possible $\Delta S=1$ sub-processes to consider are a weak
transition with emission of a pion or lepton pair and an
internal weak transition.  These are illustrated in Fig. 3 for a
three quark system. We estimate the amplitude for each of the
sub-processes and calculate the overall matrix element for
transition to the $\Lambda \Lambda$
system as a product of the sub-process amplitudes. \\

\vspace{5mm}
\centerline{\epsfig{file=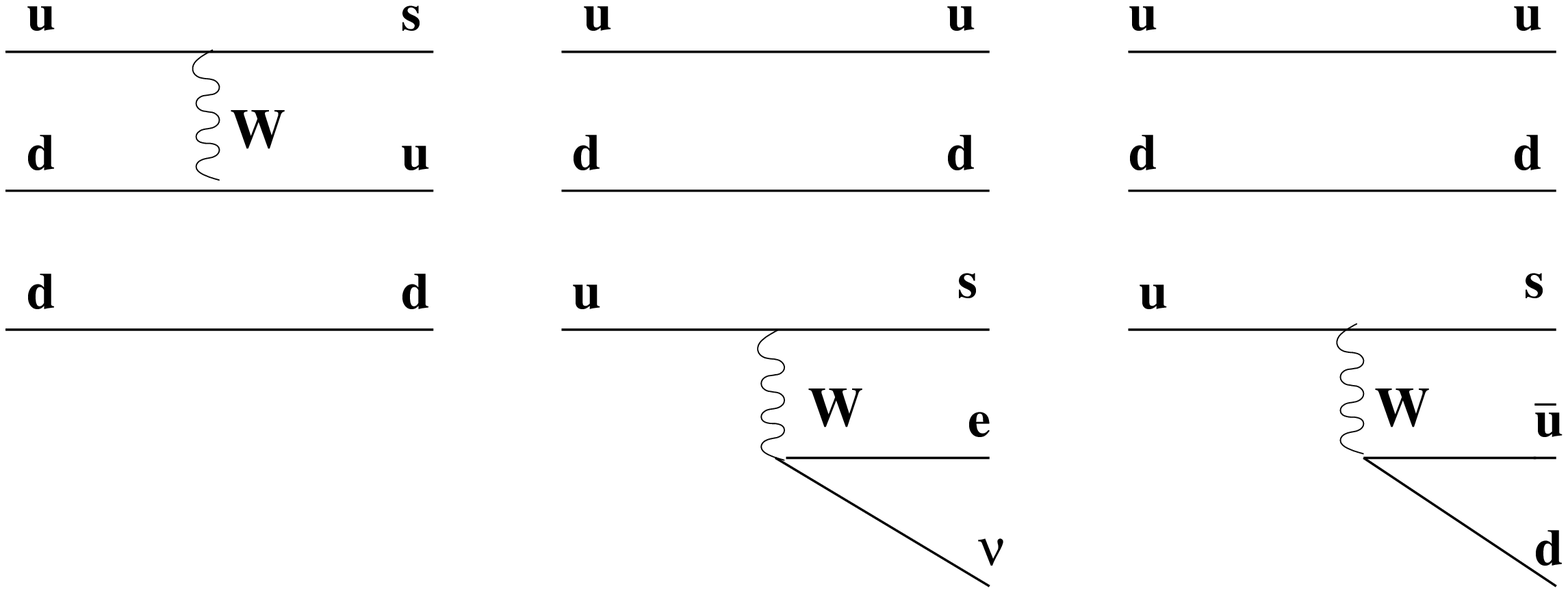,width=6cm}}{\footnotesize\textbf{
Figure 3:} Some relevant weak transitions for $NN \rightarrow HX$
}
\vspace{5mm}\\

The matrix element for weak pion emission is estimated from the
$\Lambda\rightarrow N \pi$ rate: \beq |{\cal
M}|^2_{\Lambda\rightarrow N \pi}=\frac {1}{(2\pi )^4} ~ \frac
{2m_{\Lambda} }{\Phi _2} \frac {1}{\tau _{\Lambda\rightarrow N
\pi}}=0.8 \times 10^{-12} \quad {\rm GeV}^{2}. \eeq By crossing
symmetry this is equal to the desired $|{\cal
M}|^2_{N\rightarrow \Lambda \pi}$, in the approximation of
momentum-independence which should be valid for the small
momenta in this application. Analogously, for lepton pair
emission we have \beq |{\cal M}|^2_{\Lambda\rightarrow N e\nu }=
\frac {1}{(2\pi )^4}~\frac {2 m_{\Lambda}  } {\Phi _3 }\frac
{1}{ \tau _{\Lambda\rightarrow N e\nu }} =3 \times 10^{-12}.
\eeq

The matrix element for internal conversion, $(uds) \rightarrow
(udd)$, is proportional to the spatial nucleon wave function when
two quarks are at the same point: \beq |{\cal
M}|_{\Lambda\rightarrow N} \approx <\psi _{\Lambda }|\delta^3
(\vec {r}_1-\vec {r}_2)|\psi _N > \frac {G_F \sin \theta _c \cos
\theta _c}{m_q}, \eeq where $m_q$ is the quark mass introduced in
order to make the 4 point vertex amplitude
dimensionless\cite{yaouanc}. The expectation value of the delta
function can be calculated in the harmonic oscillator model to be
\beq \label{delta1} <\psi _{\Lambda }|\delta^3 (\vec {r}_1-\vec
{r}_2)|\psi _N >~ = \left(\frac {\alpha _B}{\sqrt {2 \pi
}}\right)^3=0.4 \times 10^{-2} ~~ {\rm GeV}^3. \eeq The delta
function term can also be inferred phenomenologically in the
following way, as suggested in \cite{yaouanc}. The Fermi spin-spin
interaction has a contact character depending on $~\vec {\sigma
_1}\vec { \sigma_2}/m^2 _q \delta(\vec {r}_1-\vec {r}_2)$, and
therefore the delta function matrix element can be determined in
terms of electromagnetic or strong hyperfine splitting:
\begin{eqnarray}
(m_{\Sigma ^0}-m_{\Sigma ^+} )-(m_n-m_p)=\alpha \frac {2\pi
}{3m^2 _q}<\delta^3(\vec {r}_1-\vec {r}_2)>\\m_{\Delta} -m_N=
\alpha _S \frac {8\pi }{3 m^2 _q} <\delta^3(\vec {r}_1-\vec
{r}_2)>.
\end{eqnarray}
where $m_q$ is the quark mass, taken to be $m_N/3$. Using the
first form to avoid the issue of scale dependence of $\alpha_S$
leads to a value three times larger than predicted by the method
used in (\ref{delta1}), namely: \beq \label{delta2} <\psi
_{\Lambda }|\delta^3 (\vec {r}_1-\vec {r}_2)|\psi _N> ~ =
1.2\times 10^{-2} \quad {\rm GeV}^3. \eeq We average the
expectation values (\ref{delta1}) and (\ref{delta2}) and adopt 
\beq \label{MdeltaS}
|{\cal M}|^2_{\Lambda\rightarrow N}=4.4 \times 10^{-15}. \eeq In
this way we have roughly estimated all the matrix elements for the
relevant sub-processes based on weak-interaction phenomenology.

%\vspace{5mm}
%\centerline{\epsfig{file=NNH.eps,width=6cm}}{\footnotesize\textbf{Figure
%2:} Dominant channels: $NN \rightarrow H\pi $ and  $NN
%\rightarrow H\pi \pi$ }
%\vspace{5mm}\\

\section{Nuclear decay rates} \label{nuclifetime}

\subsection{Lifetime of doubly-strange nuclei}
\label{hypernuc}
The decay rate of a doubly-strange nucleus is:
\begin{eqnarray}
\Gamma_{A_{\Lambda \Lambda} \rightarrow A'_{H} \pi} &\approx&
K^2(2\pi )^4 \frac {m^2 _q }{2(2m_{\Lambda \Lambda})} \\
\nonumber &\times& \Phi _2 |{\cal M}|^2_{\Lambda \Lambda
\rightarrow H}.
\end{eqnarray}
where $\Phi _2$ is the two body final phase space factor, defined
as in \cite{pdb}, and $m_{\Lambda \Lambda}$ is the invariant mass
of the $\Lambda$'s, $\approx 2 m_{\Lambda}$. The factor $K$
contains the transition element in spin flavor space. It can be
estimated by counting the total number of flavor-spin states a
$uuddss$ system can occupy, and taking $K^2$ to be the fraction of
those states which have the correct quantum numbers to form the H.
That gives $K^2\sim 1/1440$, and therefore we write $K^2 = (1440~
\kappa_{1440})^{-1}$.  Combining these factors we obtain the
lifetime estimate \beq \tau_{A_{\Lambda \Lambda}\rightarrow A'_{H}
\pi}\approx \frac {3(7)~\kappa_{1440} }{|{\cal M}|^2_{\Lambda
\Lambda \rightarrow H}}10^{-18}~ {\rm s}, \eeq where the phase
space factor was evaluated for $m_H = 1.8 (2)$ GeV.

\vspace{5mm}
\centerline{\epsfig{file=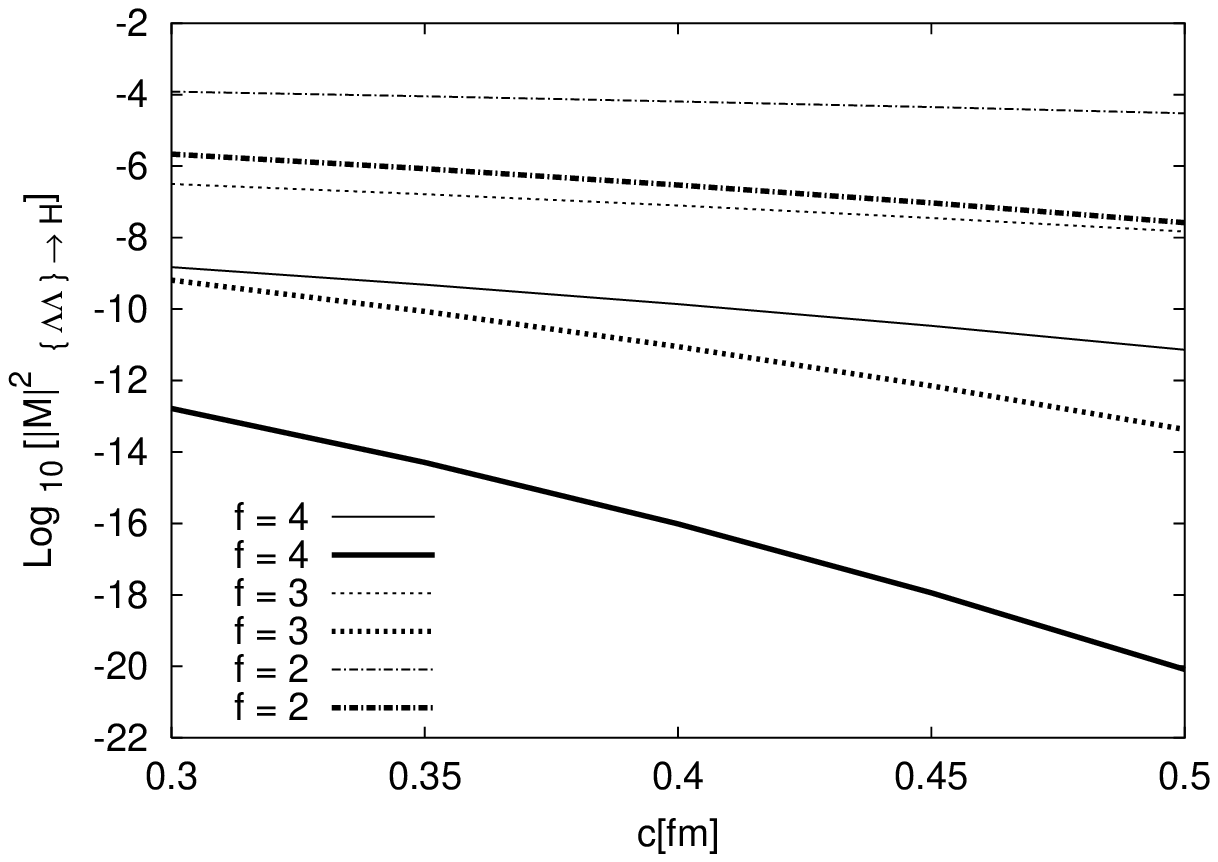,width=9cm}}{\footnotesize\textbf{
Figure 2:} Log$_{10}$ of $|{\cal M}|^2_{\Lambda \Lambda
\rightarrow H}$ versus hard core radius in fm, for f=2, 3 and 4 and two
values of the Isgur-Karl oscillator potential. Thick lines refer to $\alpha_B= 0.406~{\rm fm}^{-1}$, and thin lines to $\alpha_B = 0.221~{\rm fm}^{-1}$.}
\vspace{5mm}\\
\

Fig. 2 shows $|{\cal M}|^2_{\{\Lambda \Lambda\} \rightarrow H}$ in
the range of $f$ and hard-core radius where its value is in the
neighborhood of the experimental limits, for the standard choice
$\alpha_B = 0.406~{\rm fm}^{-1}$ and comparison value $\alpha_B =
0.221~{\rm fm}^{-1}$.  In order to suppress $\Gamma(A_{\Lambda
\Lambda}\rightarrow A'_{H} X)$ sufficiently that some $\Lambda$'s
in a double-$\Lambda$ hypernucleus will decay prior to formation
of an H, we require $|{\cal M}|^2_{\Lambda \Lambda \rightarrow H}
\lsi 10^{-8}$.  This is satisfied even for relatively large H,
e.g., $r_H \lsi r_N/2.3~(r_N/2.1) $ for a hard-core radius 0.4
(0.5) fm. Thus the observation of single $\Lambda$ decay products
from double-$\Lambda$ hypernuclei cannot be taken to exclude the
existence of an H with mass below $2 m_\Lambda$ unless it can be
demonstrated that $r_H \geq 1/2~r_N$.

%\vspace{5mm}
%\centerline{\epsfig{file=ovlaplamGZ.eps,width=8cm}}{\footnotesize\textbf{
%Figure 2:} Log$_{10}$ of $|{\cal M}|^2_{\Lambda \Lambda
%\rightarrow H}$ versus hard core radius in fm, for f=2, 3, 4.}
%\vspace{5mm}\\
%\

\subsection{Nuclear conversion to an H}
\label{convlifetimes}

If the H is actually stable ($m_H < 2 m_p + 2 m_e$) two nucleons
in a nucleus may convert to an H and cause nuclei to disintegrate.
NN $\rightarrow$ HX requires two weak reactions. Thus the rate for
the process $A_{NN}\rightarrow A'_{H}\pi \pi$, is approximately
\begin{eqnarray}
\Gamma_{A_{NN} \rightarrow  A'_{H} \pi \pi }&\approx &K^2 \frac
{(2\pi )^4} {2 (2m_{N}) }~ \Phi_3\\ \nonumber
&\times & \left(
\frac { |{\cal M}|_{N\rightarrow \Lambda \pi} ^2 |{\cal
M}|_{\Lambda \Lambda \rightarrow H} } {(2m_{\Lambda }-m_H )^2
}\right) ^2
\end{eqnarray}
where the denominator is introduced to correct the dimensions in a
way suggested by the $\Lambda \Lambda$ pole approximation. Since
other dimensional parameters relevant to this process, e.g., $m_q
= m_N/3$ or $\Lambda_{QCD}$, are comparable to $2m_{\Lambda }-m_H
$ and we are only aiming for an order-of-magnitude estimate, any
of them could equally well be used. The lifetime for nuclear
disintegration with two pion emission is thus \beq
\tau_{A_{NN}\rightarrow A'_{H}\pi \pi}\approx \frac
{40~\kappa_{1440}}{ |{\cal M}|^2_{\Lambda \Lambda \rightarrow H}}
\quad {\rm yr}, \eeq taking $m_H = 1.5$ GeV in the phase space
factor.  For the process with one pion emission and an internal
conversion, our rate estimate is
\begin{eqnarray}
\Gamma_{A_{NN}\rightarrow A'_{H}\pi}&\approx &K^2\frac {(2\pi
)^4}{2 (2m_{N})}~\Phi_2 \\ \nonumber
&\times &(|{\cal
M}|_{N\rightarrow \Lambda \pi} |{\cal M}|_{N\rightarrow \Lambda}
|{\cal M}|_{\Lambda \Lambda \rightarrow H})^2
\end{eqnarray}
leading to a lifetime, for $m_H = 1.5$ GeV, of \beq
\tau_{A_{NN}\rightarrow A'_{H} \pi}\approx \frac
{3~\kappa_{1440}}{ |{\cal M}|^2_{\Lambda \Lambda \rightarrow H}}
\quad {\rm yr}. \eeq

If $m_H \gsi 1740$ MeV, pion emission is kinematically forbidden
and the relevant final states are $e^+ \nu$ or $\gamma$; we now
calculate these rates. For the transition $A_{NN}\rightarrow A' _H
e\nu$, the rate is
\begin{eqnarray}
\Gamma_{A_{NN} \rightarrow  A'_{H}e\nu }&\approx &K^2\frac {(2\pi
)^4}{2 (2m_{N})}\Phi_3 \\ \nonumber &\times &(|{\cal
M}|_{N\rightarrow \Lambda e\nu} |{\cal M}|_{N\rightarrow \Lambda}
|{\cal M}|_{\Lambda \Lambda \rightarrow H})^2.
\end{eqnarray}
In this case, the nuclear lifetime is \beq \label{enu}
\tau_{A_{NN}\rightarrow A'_{H} e\nu}\approx \frac {\kappa_{1440}}{
|{\cal M}|^2_{\Lambda \Lambda \rightarrow H}}~10^{5} \quad {\rm
yr}, \eeq taking $m_H = 1.8$ GeV.  For $A_{NN}\rightarrow A' _H
\gamma$, the rate is approximately
\begin{eqnarray}
\Gamma_{A_{NN}\rightarrow A'_{H}\gamma }&\approx &K^2 (2\pi )^4
\frac {\alpha _{EM} m^2 _q}{2 (2m_{N})} \\ \nonumber &\times &
\Phi_2(|{\cal M}|^2 _{N\rightarrow \Lambda} |{\cal M}|_{\Lambda
\Lambda \rightarrow H})^2,
\end{eqnarray}
leading to the lifetime estimate \beq \tau_{A_{NN}\rightarrow
A'_{H} \gamma}\approx \frac {2~\kappa_{1440}}{ |{\cal
M}|^2_{\Lambda \Lambda \rightarrow H}}~10^6\quad {\rm yr}, \eeq
for $m_H = 1.8$ GeV.

One sees from Figure 1 that a lifetime bound of $\gsi {\rm
few}~10^{29}$ yr is not a very stringent constraint on this
scenario if $m_H$ is large enough that pion final states are not
allowed.  E.g., with $\kappa_{1440} = 1$ the rhs of eqn
(\ref{enu}) is $\gsi {\rm few}~10^{29}$ yr, for standard
$\alpha_B$, a hard core radius of 0.45 fm, and $r_H \approx 1/5
~r_N$ -- in the middle of the range expected based on the glueball
analogy.  If $m_H$ is light enough to permit pion production,
experimental constraints are much more powerful.  We therefore
conclude that $m_H \lsi 1740$ MeV is disfavored and is likely to
be excluded, depending on how strong limits SuperK can give.

\begin{table}[hpb]
%\small
%\tiny
\caption{The final particles and momenta for nucleon-nucleon
transitions to H in nuclei. For the 3-body final states marked
with *, the momentum given is for the configuration with H
produced at rest.} \label{t1}
\begin{center}
\begin{tabular}{|c|c|c|c|}
\hline

mass        & final state & final momenta  & partial lifetime  \\
$m_H$ [GeV] & A$^\prime$ H +    & p [MeV]        & $ \times
K^2|{\cal M}|^2 _{\Lambda \Lambda \rightarrow H}$ [yr] \\ \hline

1.5         &  $\pi $     & 318            & $2~10^{-3}$    \\ \hline
1.5         &  $\pi \pi$  & 170*           & 0.03           \\ \hline
1.8         &  $e \nu$    & 48*            & 70             \\ \hline
1.8         &  $\gamma$   & 96             & $2~10^3$       \\ \hline 

\end{tabular}
\end{center}
\end{table}

\section{Decays of a Quasi-stable H} \label{metastable}

If $2 m_N \lsi m_H < m_N + m_\Lambda$, the H is not stable but it proves to be very long lived if its wavefunction is compact
enough to satisfy the constraints from doubly-strange hypernuclei
discussed in sections \ref{expts} and \ref{hypernuc}.  The limits
on nuclear stability discussed in the previous section do not
apply here because nuclear disintegration to an H is not
kinematically allowed.  

\subsection{Wavefunction Overlap}

To calculate the decay rate of the H we start from the transition
matrix element (\ref{matrixel}). In contrast to the calculation of
nuclear conversion rates, the outgoing nucleons are asymptotically
plane waves. Nonetheless, at short distances their repulsive
interaction suppresses the relative wavefunction at short
distances much as in a nucleus.  It is instructive to compute the
transition amplitude using two different approximations.  First, we treat the nucleons
%(equivalent to $ \Lambda$'s, apart from the weak interactions) 
as plane waves so the spatial amplitude is:
\begin{eqnarray}
T_{H\rightarrow \Lambda \Lambda }&=&2 \pi i\delta (E_{\Lambda
\Lambda}-E_H) \int \prod _{i=a,b}
d^3 \rho^i d^3 \lambda ^i d^3 a~ d^3 R_{CM} \nonumber \\
&\times & \psi _H \psi ^{*a} _{\Lambda}~\psi^{*b} _{\Lambda}~
e^{i({\vec k}^a _N+{\vec k}^b _N-{\vec k}_{H}){\vec R}_{CM}}.
\end{eqnarray}
The integration over ${\vec R}_{CM}$ gives the usual 4D $\delta$
function. After performing the remaining integrations leading to
$|{\cal M}|_{H\rightarrow \Lambda \Lambda}$, as in
(\ref{matrixel2}), the amplitude is:
\begin{eqnarray}
\label{planewaves}
|{\cal M}|_{H \rightarrow \Lambda
\Lambda}&=&\left (\frac {2f}{1+f^2}\right )^6 \left( \frac{3}{2}
\right)^{3/4}\left( \frac{\alpha _H}{\sqrt{\pi}} \right)^{3/2}\\
\nonumber &\times & \int^{\infty} _{0} d^3 a~ e ^{-\frac
{3}{4}\alpha_{H} ^2 a^2-i\frac{{\vec k}^a _N-{\vec k}^b
_N}{2}{\vec a}}\\ \nonumber &=& \left( \frac{8}{3\pi}
\right)^{3/4} \left (\frac {2f}{1+f^2}\right )^6 \alpha ^{-3/2}
_H~e^{-\frac{({\vec k}^a _N-{\vec k}^b _N)^2}{12~\alpha ^2 _H}}.
\end{eqnarray}
The amplitude depends on the size of the H through the factor $ f
= r_N/r_H$.  Note that the normalization $N_{BG}$ in the analogous
result (\ref{overlap}) which comes from the Bethe-Goldstone
wavefunction of $\Lambda$'s in a nucleus has been replaced in this
calculation by the plane wave normalization factor $1/\sqrt{V}$
which cancels with the volume factors in the phase space when
calculating transition rates.

Transition rates calculated using eq. (\ref{planewaves}) provide
an upper limit on the true rates, because the calculation neglects
the repulsion of two nucleons at small distances. To estimate the
effect of the repulsion between nucleons we again use the
Bethe-Goldstone solution with the hard core potential. It has the
desired properties of vanishing inside the hard core radius and
rapidly approaching the plane wave solution away from the hard
core.  As noted in section \ref{BBG}, $N_{BG}\rightarrow
1/\sqrt{V}$, for $a\rightarrow \infty$. Therefore, we can write
the transition amplitude as in (\ref {overlap}), with the
normalization factor $1/\sqrt{V}$ canceled with the phase space
volume element:
\begin{eqnarray}
|{\cal M}|_{H \rightarrow \Lambda \Lambda }&=&\left
(\frac {2f}{1+f^2}\right )^6 \left( \frac{3}{2}
\right)^{3/4}\left( \frac{\alpha _H}{\sqrt{\pi}} \right)^{3/2}\\
\nonumber &\times & \int^{\infty} _{0} d^3 a \frac {u(k_F a)}{k_F
a}e ^{-\frac {3}{4}\alpha_{H} ^2 a^2}.
\end{eqnarray}
This should give the most realistic estimate of decay rates.  Table 2
shows the overlap values for a variety of choices of $r_H$,
hard-core radii, and $\alpha_B$.

\begin{table}[hpb]
%\small
%\tiny
\caption{$|{\cal M}|_{H \rightarrow \Lambda \Lambda }^2 $ in ${\rm
GeV}^{-3/2}$ for different values of $1/f$ (rows) and
nuclear hard core (columns), for $\alpha_{B1}=0.406~{\rm fm}^{-1}$ and $\alpha_{B2}=0.221~{\rm fm}^{-1}$ .}\label{t2}
\begin{center}
\begin{tabular}{|c|c|c|c|c|}
\hline

  &  \multicolumn{2}{c|} {0.4 fm}    & \multicolumn{2}{c|} {0.5 fm}               \\\cline{2-5} 
            & $\alpha_{B1}$ & $\alpha_{B2}$ & $\alpha_{B1}$ & $\alpha_{B2}$ \\ \hline 
0.3         & $~2~10^{-10}~$ & $~4~10^{-6}~$     &  $~4~10^{-13}~$ & $~6~10^{-7}~$      \\ \hline
0.4         & $~9~10^{-7}~$ & $8~10^{-4}$     & $~2~10^{-8}~$ & $~3~10^{-4}~$           \\ \hline
0.5         & $~4~10^{-4}~$ & $~0.02~$           & $~1~10^{-5}~$ & $~0.01~$             \\ \hline

\end{tabular}
\end{center}
\end{table}

\subsection{ Decay rates and lifetimes}

Starting from $|{\cal M}|_{H \rightarrow \Lambda \Lambda}$ we can
calculate the rates for H decay in various channels, as we did for nuclear
conversion in the previous section. The rate of $H\rightarrow nn$
decay is
\begin{eqnarray}\label{HtoNN}
\Gamma_{H\rightarrow nn}&\approx &K^2\frac {(2\pi )^4m^5 _q}{2~
m_{H}}~\Phi_2 (m_H)\\ \nonumber &\times & ( |{\cal M}|^2
_{N\rightarrow \Lambda} |{\cal M}|_{H \rightarrow \Lambda
\Lambda})^2.
\end{eqnarray}
The lifetime for this transition, for $m_H = 1.9 ~(2)$ GeV, is
\beq \label{tau2wk} \tau_{H\rightarrow NN} \approx 5(2)~10^{12} ~\kappa
_{1440}~\mu_0 \quad {\rm yr}, \eeq where we have introduced
$|{\cal M}|_{H \rightarrow \Lambda \Lambda}^2 \equiv
7~10^{-8}/\mu_0$; values of $\mu_0 \ge 1$ are consistent with
hypernuclear constraints.  Thus we see that the H is stable on cosmological time scales if its mass is $\lsi 2.04$ GeV.

If $2.04~ {\rm GeV} < m_H < 2.23$ GeV, H decay requires only a single weak interaction, so the rate in eq.  (\ref{HtoNN}) must be divided by $ |{\cal M}|^2
_{N\rightarrow \Lambda}$ given in eqn (\ref{MdeltaS}).  Thus we have \beq \label{tau1wk} \tau_{H\rightarrow N \Lambda} \approx 10^5 ~{\rm sec}~~\kappa
_{1440}~\mu_0. \eeq Finally, if $ m_H > 2.23$ GeV, there is no weak interaction suppression and
\beq \label{tau0wk} \tau_{H\rightarrow \Lambda \Lambda} \approx 10^{-9} ~\kappa
_{1440}~\mu_0 ~{\rm sec}.\eeq 
Equations (\ref{tau2wk})-(\ref{tau0wk}) with  $\mu_0 = 1$ give the lower bound on the H lifetime, depending on its mass. 

Our results for the H lifetime are dramatically different from the classic calculation of Donoghue, Golowich, and Holstein \cite{donoghue:Hlifetime}, because we rely on experiment to put an upper limit on the wavefunction overlap $|{\cal M}|_{H \rightarrow \Lambda \Lambda}^2$. The bag model is not a particularly good description of sizes of hadrons, and in the treatment of \cite{donoghue:Hlifetime} the H size appears to be fixed implicitly to some value which may not be physically realistic. Furthermore, it is hard to tell whether the bag model analysis gives a good accounting of the known hard core repulsion between nucleons.   As our calculation of previous sections shows, these are crucial parameters in determining the overlap.   Our treatment of the color-flavor-spin and weak interaction parts of the matrix elements is rough but should give the correct order-of-magnitude, so the difference in lifetime predictions of the two models indicates that the spatial overlap in the bag model is far larger than in our model using a standard hard core and taking $r_H \approx 0.4$ fm (which accounts for the  suppression measured in hypernuclear experiments).  The calculation of the weak interaction and color-flavor-spin matrix elements in ref. \cite{donoghue:Hlifetime} is more detailed than ours and should be more accurate.  It could be combined with a phenomenological approach to the spatial wavefunction overlap to provide a more accurate yet more general analysis.  We note that due to the small size of the H, the p-wave contribution should be negligible.  

One would like to use experiment to place limits on the product of the
local density of H's and the H decay rate, if the H is long-lived enough to be dark matter, i.e., $m_H < m_N + m_\Lambda$.
Estimates of the local number density of H's at various depths in the Earth, assuming the
dark matter consists of H's and $\bar{\rm H}$'s, will be discussed
in ref. \cite{fz:lims}.  The Sudbury Neutrino Observatory (SNO) can probably place good limits on the rate of $H \rightarrow nn$ in that detector.  
The next most important channel $H\rightarrow nn \gamma $ should
be easy to detect in SuperK for photon energy in the low-background range
$\approx 20-100$ MeV \cite{SKgamma}, or in Kamland for lower photon energies.\footnote{GRF thanks T. Kajita for discussions on these issues.} The rate is:
\begin{eqnarray}
\Gamma_{H\rightarrow nn\gamma}&\approx &K^2\alpha _{EM} \frac
{(2\pi )^4m^3 _q}{2~ m_{H}}~\Phi_3 (m_H)\\ \nonumber &\times & (
|{\cal M}|^2 _{N\rightarrow \Lambda} |{\cal M}|_{H\rightarrow
\Lambda \Lambda})^2
\end{eqnarray}
leading to \beq \tau_{H\rightarrow NN\gamma} \approx
2~10^{19}~(3~10^{17})~\kappa _{1440}~\mu_0 \quad {\rm yr}, \eeq
for $m_H= 1.9~(2)~{\rm GeV}$. The lifetime for $H\rightarrow
npe\nu$ is similar in magnitude.  It is more sensitive to $m_H$
due to the 4-body phase space.
\begin{eqnarray}
\Gamma_{H\rightarrow pne\nu}&\approx &K^2\frac {(2\pi )^4~m_q}{2~
m_{H}}~\Phi_4 (m_H)\\ \nonumber &\times & ( |{\cal M}|
_{N\rightarrow \Lambda}|{\cal M}| _{N\rightarrow \Lambda e\nu}
|{\cal M}|_{H\rightarrow \Lambda \Lambda})^2
\end{eqnarray}
and \beq \tau_{H\rightarrow pne\nu} \approx
5~10^{19}~(3~10^{16})~\kappa _{1440}~\mu_0 \quad {\rm yr}. \eeq
for $m_H= 1.9~(2)~{\rm GeV}$ .

\section{Summary} \label{summary}

We have considered the constraints placed on the H di-baryon by
the stability of nuclei and hypernuclei with respect to conversion
to an H dibaryon, and we have calculated the lifetime of the H if
it is not stable. We used the Isgur-Karl wavefunctions for quarks
in baryons and the H, and the Bethe-Goldstone wavefunction for
nucleons in a nucleus, to obtain a rough estimate of the
H-baryon-baryon wavefunction overlap.  Observation of $\Lambda$
decays in double-$\Lambda$ hypernuclei is shown not to exclude an
H as long as $r_H \lsi 1/2~ r_N$.

Combining our wavefunction overlap estimates with phenomenological
weak interaction matrix elements, permits the lifetime of the H and the rate for
conversion of nuclei to H to be estimated.  The results depend radically on which channels are kinematically allowed, and hence on the H mass, since for each weak interaction required there is a substantial suppression in the matrix element.  Our estimates have uncertainties of greater than an order of magnitude: the weak
interaction matrix elements are uncertain to a factor of a few,
factors of order 1 were ignored, a crude statistical estimate for
the flavor-spin overlap was used, mass scales were set to $m_N/3$,
and, most importantly, the models used to calculate the
wavefunction overlap surely oversimplify the physics. The wavefunction overlap
is highly uncertain because it depends on nuclear wavefunctions
and hadronic dynamics which are not adequately understood at
present.  Nonetheless, the enormous suppression of H production
and decay rates found in the model calculation means that the observation of double-$\Lambda$ hypernuclei does not exclude $m_H < 2 m_\Lambda$.  It is even conceivable that an absolutely
stable H is possible\cite{f:stableH}.

We calculate the lifetime of the H for various mass ranges, taking the H wavefunction to be compact enough that hypernuclear constraints are satisfied. If the H decays through strong interactions, $m_H > 2 m_\Lambda$, its lifetime is $\gsi 10^{-9}$ sec.  If its  mass is in the range $m_N + m_\Lambda \lsi m_H
\lsi 2 m_\Lambda$, its lifetime is longer than a few $10^5$ sec, and if  $2 m_N \lsi m_H
\lsi m_N + m_\Lambda$ the H lifetime is at least an order of magnitude greater than the lifetime of the Universe.  Note that these lifetime bounds do not suffer from the large uncertainties associated with estimating the wavefunction overlap because, for a given value of $m_H$, the H lifetime can be related to the hypernuclear conversion rate for which we have a rough experimental upper limit.  It may be that the production rate of H's in double $\Lambda$ hypernuclei is about equal to the $\Lambda$ decay rate, so that an appreciable fraction of double $\Lambda$ hypernuclei produce an H.  One might hope that in this case the H could be observed by reconstructing it through its decay products, e.g., $H \rightarrow \Sigma^- p$.  Unfortunately, however, the long lifetimes implied by the limit on the wavefunction overlap mean that the H's would diffuse out of the apparatus before decaying.\footnote{GRF thanks K. Imai for enlightening discussions of this topic.}  

SuperK can place important constraints on the conjecture of an
absolutely stable H, or conceivably discover evidence of its
existence, through observation of the pion(s), positron, or photon
produced when two nucleons in an oxygen nucleus convert to an H.
We estimate that SuperK could achieve a lifetime limit $\tau \gsi
{\rm few} ~10^{29}$ yr which is in the estimated lifetime range for
$m_H \gsi 1740$ MeV and $r_H \approx 1/5 ~r_N$. SuperK and SNO can also place limits on signatures of H 
decays if it is not absolutely stable yet contributes to the dark matter of the Universe. This calculation will be reported
elsewhere. The possibility that H and anti-H were produced in
sufficient abundance in the early universe to account for the dark
matter and baryon asymmetry will also be elaborated elsewhere.

The research of GRF was supported in part by NSF-PHY-0101738. GRF
acknowledges helpful conversations with many colleagues,
particularly G. Baym, A. Bondar, T. Kajita, M. May, M.
Ramsey-Musolf, and P. Vogel.  GZ wishes to thank Allen Mincer and
Marko Kolanovic for useful advice and is grateful to Emiliano
Sefusatti for many helpful comments.

%\bibliographystyle{unsrt}
%\bibliography{stabilityH}

\end{document}